\documentclass[smallextended]{svjour3}
\smartqed 
\usepackage{graphicx}
\usepackage[table,xcdraw]{xcolor}
\usepackage{authblk}
\usepackage[utf8]{inputenc}
\usepackage{color}
\usepackage{graphicx}
\usepackage{amsmath}
\usepackage{amssymb}
\usepackage[colorlinks=true, citecolor=blue]{hyperref}
\usepackage{multicol}

\journalname{Experimental Astronomy}

\begin{document}

\title{The missing link in gravitational-wave astronomy}
\subtitle{A summary of discoveries waiting in the decihertz range}

\author{Manuel~Arca~Sedda \and
Christopher~P~L~Berry \and
Karan~Jani \and
Pau~Amaro-Seoane \and
Pierre~Auclair \and
Jonathon~Baird \and
Tessa~Baker \and
Emanuele~Berti \and
Katelyn~Breivik \and
Chiara~Caprini \and
Xian~Chen \and
Daniela~Doneva \and
Jose~M~Ezquiaga \and
K~E~Saavik~Ford \and
Michael~L~Katz \and
Shimon~Kolkowitz \and
Barry~McKernan \and
Guido~Mueller \and
Germano~Nardini \and
Igor~Pikovski \and
Surjeet~Rajendran \and
Alberto~Sesana \and
Lijing~Shao \and
Nicola~Tamanini \and
Niels~Warburton \and
Helvi~Witek \and
Kaze~Wong \and
Michael~Zevin
}

\date{Received: 3 August 2020 / Accepted: 9 February 2021 / Published online: 29 April 2021}

\institute{M Arca Sedda \at Astronomisches Rechen-Institut, Zentr\"um f\"ur Astronomie, Universit\"at Heidelberg, M\"onchofstr. 12-14, Heidelberg, Germany \and
C P L Berry \at Center for Interdisciplinary Exploration and Research in Astrophysics (CIERA), Department of Physics and Astronomy, Northwestern University, 2145 Sheridan Road, Evanston, IL 60208, USA\\
SUPA, School of Physics and Astronomy, University of Glasgow, Glasgow G12 8QQ, UK
\email{christopher.berry@northwestern.edu} \and
K Jani \at Department of Physics and Astronomy, Vanderbilt University, Nashville, TN 37212, USA \and
P Amaro-Seoane \at
Universitat Polit{\`e}cnica de Val{\`e}ncia, IGIC, Spain\\
Kavli Institute for Astronomy and Astrophysics, Beijing 100871, China\\
Institute of Applied Mathematics, Academy of Mathematics and Systems Science, CAS, Beijing 100190, China\\
Zentrum f{\"u}r Astronomie und Astrophysik, TU Berlin, Hardenbergstra{\ss}e 36, 10623 Berlin, Germany \and
P Auclair \at Laboratoire Astroparticule et Cosmologie, CNRS UMR 7164, Universit\'{e} Paris-Diderot, 10 rue Alice Domon et L\'{e}onie Duquet, 75013 Paris, France \and
J Baird \at High Energy Physics Group, Physics Department, Imperial College London, Blackett
Laboratory, Prince Consort Road, London, SW7 2BW, UK \and
T Baker \at School of Physics and Astronomy, Queen Mary University of London, Mile End Road, London, E1 4NS, UK \and
E Berti \at Department of Physics and Astronomy, Johns Hopkins University, 3400 N.\ Charles Street, Baltimore, MD 21218, USA \and
Katelyn Breivik \at Canadian Institute for Theoretical Astrophysics, University of Toronto, 60 St.\ George Street, Toronto, Ontario, M5S 1A7, Canada \and
C Caprini \at Laboratoire Astroparticule et Cosmologie, CNRS UMR 7164, Universit\'{e} Paris-Diderot, 10 rue Alice Domon et L\'{e}onie Duquet, 75013 Paris, France \and
X Chen \at Astronomy Department, School of Physics, Peking University, Beijing 100871, China \\
Kavli Institute for Astronomy and Astrophysics, Peking University, Beijing 100871, China \and 
D Doneva \at Theoretical Astrophysics, Eberhard Karls University of T\"ubingen, T\"ubingen 72076, Germany \and 
J M Ezquiaga \at Kavli Institute for Cosmological Physics, Enrico Fermi Institute, The University of Chicago, Chicago, IL 60637, USA \and
K E S Ford \at City University of New York-BMCC, 199 Chambers St, New York, NY 10007, USA\\
Department of Astrophysics, American Museum of Natural History, New York, NY 10028, USA \and
M L Katz \at Center for Interdisciplinary Exploration and Research in Astrophysics (CIERA), Department of Physics and Astronomy, Northwestern University, 2145 Sheridan Road, Evanston, IL 60208, USA \and
S Kolkowitz \at Department of Physics, University of Wisconsin -- Madison, Madison, WI 53706, USA \and
B McKernan \at City University of New York-BMCC, 199 Chambers St, New York, NY 10007, USA\\
Department of Astrophysics, American Museum of Natural History, New York, NY 10028, USA \and
G Mueller \at Department of Physics, University of Florida, PO Box 118440, Gainesville, Florida 32611, USA \and
G Nardini \at Faculty of Science and Technology, University of Stavanger, 4036 Stavanger, Norway \and
Igor Pikovski \at Department of Physics, Stevens Institute of Technology, Hoboken, NJ 07030, USA\\
Department of Physics, Stockholm University, SE-10691 Stockholm, Sweden \and
S Rajendran \at Department of Physics and Astronomy, Johns Hopkins University, 3400 N. Charles Street, Baltimore, MD 21218, USA \and
A Sesana \at Universit\`a di Milano Bicocca, Dipartimento di Fisica G.\ Occhialini, Piazza della Scienza 3, I-20126, Milano, Italy \and
L Shao \at Kavli Institute for Astronomy and Astrophysics, Peking University, Beijing 100871, China \\
National Astronomical Observatories, Chinese Academy of Sciences, Beijing 100012, China \and
N Tamanini \at Max-Planck-Institut f\"ur Gravitationsphysik (Albert-Einstein-Institut), Am Mühlenberg 1, 14476 Potsdam-Golm, Germany \and 
N Warburton \at School of Mathematics and Statistics, University College Dublin, Belfield, Dublin 4, Ireland \and
H Witek \at Department of Physics, King's College London, Strand, London, WC2R 2LS, UK \and 
K Wong \at Department of Physics and Astronomy, Johns Hopkins University, 3400 N. Charles Street, Baltimore, MD 21218, USA \and
M Zevin \at Center for Interdisciplinary Exploration and Research in Astrophysics (CIERA), Department of Physics and Astronomy, Northwestern University, 2145 Sheridan Road, Evanston, IL 60208, USA
}

\authorrunning{Arca~Sedda, Berry, Jani \textit{et al}.}

\maketitle

\begin{abstract}
    Since 2015 the gravitational-wave observations of LIGO and Virgo have transformed our understanding of compact-object binaries. 
    In the years to come, ground-based gravitational-wave observatories such as LIGO, Virgo, and their successors will increase in sensitivity, discovering thousands of stellar-mass binaries. 
    In the 2030s, the space-based \textit{LISA} will provide gravitational-wave observations of massive black holes binaries. 
    Between the $\sim10$--$10^3~\mathrm{Hz}$ band of ground-based observatories and the $\sim10^{-4}$--$10^{-1}~\mathrm{Hz}$ band of \textit{LISA} lies the uncharted decihertz gravitational-wave band. 
    We propose a \emph{Decihertz Observatory} to study this frequency range, and to complement observations made by other detectors. 
    Decihertz observatories are well suited to observation of intermediate-mass ($\sim10^2$--$10^4 M_\odot$) black holes; they will be able to detect stellar-mass binaries days to years before they merge, providing early warning of nearby binary neutron star mergers and measurements of the eccentricity of binary black holes, and they will enable new tests of general relativity and the Standard Model of particle physics.
    Here we summarise how a Decihertz Observatory could provide unique insights into how black holes form and evolve across cosmic time, improve prospects for both multimessenger astronomy and multiband gravitational-wave astronomy, and enable new probes of gravity, particle physics and cosmology.
    
\end{abstract}

\section{The gravitational-wave spectrum}

When new frequency ranges of the electromagnetic spectrum became open to astronomy, our understanding of the Universe expanded as we gained fresh insights and discovered new phenomena~\cite{Longair:2006}. 
Equivalent breakthroughs are awaiting gravitational-wave (GW) astronomy~\cite{Sathyaprakash:2009xs,Abbott:2016blz}.
Here, we summarise the \emph{scientific potential of exploring the ${\sim0.01}$--${1~\mathrm{{Hz}}}$ GW spectrum}.

The first observation of a GW signal was made in 2015 by the Laser Interferometer Gravitational-Wave Observatory (LIGO)~\cite{Abbott:2016blz}. 
Ground-based detectors such as LIGO~\cite{TheLIGOScientific:2014jea}, Virgo~\cite{TheVirgo:2014hva}, and KAGRA~\cite{Akutsu:2018axf} observe over a frequency spectrum $\sim10$--$10^3~\mathrm{Hz}$. 
This is well tailored to the detection of merging stellar-mass black hole (BH) and neutron star (NS) binaries~\cite{LIGOScientific:2018mvr}. 
Next-generation ground-based detectors, like Cosmic Explorer~\cite{Evans:2016mbw} or the Einstein Telescope~\cite{Sathyaprakash:2012jk,Hild:2010id} may observe down to a few hertz. 
Only a small part of the GW spectrum can thus be observed by ground-based detectors, and extending to lower frequencies requires space-based observatories.

Lower frequency GW signals originate from coalescences of more massive binaries, and stellar-mass binaries earlier in their inspirals. 
Due for launch in 2034, the \textit{Laser Interferometer Space Antenna} (\textit{LISA}) will observe across frequencies $\sim10^{-4}$--$10^{-1}~\mathrm{Hz}$~\cite{Audley:2017drz}, 
optimal for mergers of binaries with $\sim10^6 M_\odot$ massive BHs~\cite{Klein:2015hvg,Babak:2017tow,Berry:2019wgg}. 
\textit{LISA} will be able to observe nearby stellar-mass binary BHs (BBHs) years--days prior to merger~\cite{Sesana:2016ljz}, when they could be observed by ground-based detectors. 
\emph{Multiband} observations of BBHs would provide improved measurements of source properties~\cite{Vitale:2016rfr,Jani:2019ffg,Liu:2020nwz}, new constraints on their formation channels~\cite{Breivik:2016ddj,Nishizawa:2016jji}, and enable precision tests of general relativity (GR)~\cite{Vitale:2016rfr,Toubiana:2020vtf}.

Pulsar timing arrays are sensitive to even lower frequency GWs of $\sim10^{-9}$--$10^{-7}~\mathrm{Hz}$~\cite{IPTA):2013lea}, 
permitting observation of $\sim10^9 M_\odot$ supermassive BHs~\cite{Mingarelli:2017fbe}. 
Combining \textit{LISA} and pulsar timing observations will produce new insights into the evolution of (super)massive BHs~\cite{Pitkin:2008iu,Colpi:2019yzd}.

The case for extending the accessible GW spectrum with an observatory that can observe in the $\sim 0.01$--$1~\mathrm{Hz}$ \emph{decihertz} range is explained in \cite{Sedda:2019uro}, based upon a White Paper that we submitted in response to ESA's Voyage 2050 call, and here we summarise the highlights. 
Decihertz observations would:
\begin{enumerate}
    \item Reveal the formation channels of stellar-mass binaries, complementing ground-based observations with deep multiband observations. 
    \item Complete our census of the population of BHs by enabling unrivaled measurements of intermediate-mass BHs (IMBHs), which may be the missing step in the evolution of (super)massive BHs.
    \item Provide a new laboratory for tests of fundamental physics. 
\end{enumerate}
\emph{Decihertz observatories (DOs) have the capability to resolve outstanding questions about the intricate physics of binary stellar evolution, the formation of astrophysical BHs at all scales across cosmic time, and whether extensions to GR or the Standard Model of particle physics are required}.

\section{The potential of decihertz observatories}

Decihertz observations would bridge space-based low-frequency detectors and ground-based detectors, giving us access to a wide variety of astrophysical systems:
\begin{enumerate}
    \item \emph{Stellar-mass binaries comprised of compact stellar objects---white dwarfs (WDs), NSs, and stellar-mass BHs.} 
    Since BH and NS mergers are observable with ground-based detectors, a DO would allow multiband observations of these populations. 
    WDs are inaccessible to ground-based detectors~\cite{Littenberg:2019grv}, and so can only be studied with space-based detectors. 
    While the current-generation of ground-based detectors will detect stellar-mass BBHs to redshifts $z \sim 1$--$2$, next-generation detectors will discover them out to $z \sim 20$, enabling them to chart the evolution of the binary population across the history of the Universe~\cite{Kalogera:2019sui}; a DO could match this range, far surpassing \textit{LISA}. 
    Furthermore, decihertz observations of compact-object mergers would provide valuable forewarning of multimessenger emission associated with merger events. 
    If detected, multimessenger observations reveal details about the equation of state of nuclear density matter~\cite{Abbott:2018exr,Montana:2018bkb,Most:2018hfd,Coughlin:2018fis,Margalit:2019dpi}, the production of heavy elements~\cite{Abbott:2017wuw,Chornock:2017sdf,Tanvir:2017pws,Wanajo:2018wra,Siegel:2018zxq}, and provide a unique laboratory for testing gravity~\cite{Monitor:2017mdv,Abbott:2018lct,Belgacem:2018lbp,Belgacem:2019pkk}, as well as potentially identifying the progenitors of Type Ia supernovae~\cite{Hillebrandt:2013gna,Maoz:2013hna,Mandel:2017pzd}. 
    Even without finding a counterpart, correlation with galaxy catalogues can provide \emph{standard siren} cosmological measurements~\cite{Schutz:1986gp,MacLeod:2007jd,Chen:2017rfc,Abbott:2019yzh,Kyutoku:2016zxn,DelPozzo:2017kme,Cutler:2009qv,Nishizawa:2010xx}. 
    Following their detection by LIGO and Virgo, BHs and NSs are a \emph{guaranteed} class of GW source~\cite{Abbott:2016blz,LIGOScientific:2018mvr,Abbott:2020niy}.
    With a large number of observations, we can infer the formation channels for compact-object binaries, and the physics that governs them~\cite{Mandel:2009nx,Stevenson:2017dlk,Talbot:2017yur,Zevin:2017evb,Barrett:2017fcw,Sedda:2018nxm,Kalogera:2019sui,Sedda:2020vwo,Farmer:2020xne}. 
    Eccentricity is a strong indicator of formation mechanism~\cite{Nishizawa:2016jji,Breivik:2016ddj,Nishizawa:2016eza,Canuel:2017rrp,Kremer:2018tzm,Randall:2019znp}; however, residual eccentricity is expected to be small in the regime observable with ground-based detectors~\cite{Peters:1964zz,TheLIGOScientific:2016htt,Samsing:2017rat,Rodriguez:2018pss} while in some cases, BBHs formed with the highest eccentricities will emit GWs of frequencies too high for \textit{LISA}~\cite{Randall:2018lnh,DOrazio:2018jnv,Kremer:2018tzm,Arca-Sedda:2018qgq,Kremer:2018cir,Zevin:2018kzq,Chen:2017gfm}. 
    Hence DOs could provide unique insights into binary evolution.
    \item \emph{IMBHs of $\sim10^2$--$10^4 M_\odot$.} 
    IMBHs could be formed via repeated mergers of stars and compact stellar remnants in dense star clusters~\cite{PortegiesZwart:1999nm,Giersz:2015,Sedda:2019rfd,Abbott:2020mjq}.
    Using GWs, IMBHs could be observed in a binary with a compact stellar remnant as an intermediate mass-ratio inspiral (IMRI)~\cite{AmaroSeoane:2007aw,Brown:2006pj,Rodriguez:2011aa,Haster:2015cnn}, or in a coalescing binary with another IMBH. 
    A DO would enhance the prospects of IMRI detection to tens of events per year, with observations extending out to high redshift.
    Mergers involving a WD or a NS can lead to tidal disruption events with a bright electromagnetic counterpart~\cite{Chen:2018foj,Eracleous:2019bal}. 
    IMBHs binaries could be studied across the entire history of the Universe, 
    charting the properties of this population and constraining the upper end of the pair-instability mass gap~\cite{Ezquiaga:2020tns}, while also providing a detailed picture of the connection (or lack thereof) between IMBHs and the seeds of massive black holes~\cite{Volonteri:2009vh}. 
    The connection between massive BHs and their lower-mass counterparts could be further explored through observations of binaries orbiting massive BHs in galactic centres, or around IMBHs in smaller clusters~\cite{McKernan:2014oxa,Bartos:2016dgn,Stone:2016wzz,McKernan:2017umu,Chen:2017gfm,Chen:2017xbi,Gondan:2017wzd,Secunda:2018kar,Rasskazov:2019gjw,Yang:2019okq}. 
    BBH--IMBH systems are a target for DO--ground-based multiband observation because they emit both ${1}$--${10^2~\mathrm{{Hz}}}$ GWs and \emph{simultaneously} ${0.01}$--${1~\mathrm{{Hz}}}$ GWs.
    \item \emph{Cosmological sources as part of a stochastic GW background (SGWB).} 
    Both this and the other astrophysical sources serve as probes of new physics, enabling tests of deviations from GR and the Standard Model. 
    A first-order phase transition in the early Universe can generate a SGWB~\cite{Kosowsky:1992vn,Kamionkowski:1993fg,Gogoberidze:2007an,Caprini:2009yp,Hindmarsh:2013xza,Hindmarsh:2015qta}; a DO would be sensitive to first-order phase transitions occurring at higher temperature, or with a shorter duration, compared to \textit{LISA}. 
    A DO would be sensitive to a SGWB from a source at $\sim 1~\mathrm{TeV}$ and beyond: $\mathrm{TeV}$-scale phenomena have been consider to resolve with the hierarchy problem or the question of dark matter~\cite{Randall:2006py,Nardini:2007me,Konstandin:2010cd,Konstandin:2011dr,Bruggisser:2018mus,Megias:2018sxv}, while $100~\mathrm{TeV}$-scale phenomena appear in new solutions to the hierarchy problem such as the relaxion~\cite{Arkani-Hamed:2015vfh,Graham:2015ifn}. 
    Furthermore, SGWB (non-)detection could constrain the properties of cosmic strings~\cite{Vachaspati:1984gt,Blanco-Pillado:2013qja} down to string tensions of $\sim 10^{-19}$, while \textit{LISA} would reach $\sim 10^{-17}$~\cite{Auclair:2019wcv} and pulsar timing array observations currently constrain tensions to be $\lesssim 10^{-11}$~\cite{Sanidas:2012ee,Blanco-Pillado:2017rnf}.
    
\end{enumerate}
Decihertz observations provide a unique insight into the physics of each of these sources, and observations would answer questions on diverse topics ranging from the dynamics of globular clusters to the nature of dark matter.

\section{Decihertz mission concepts}

\begin{figure}
\centering
 \includegraphics[scale=0.38, trim = {80 60 100 80}]{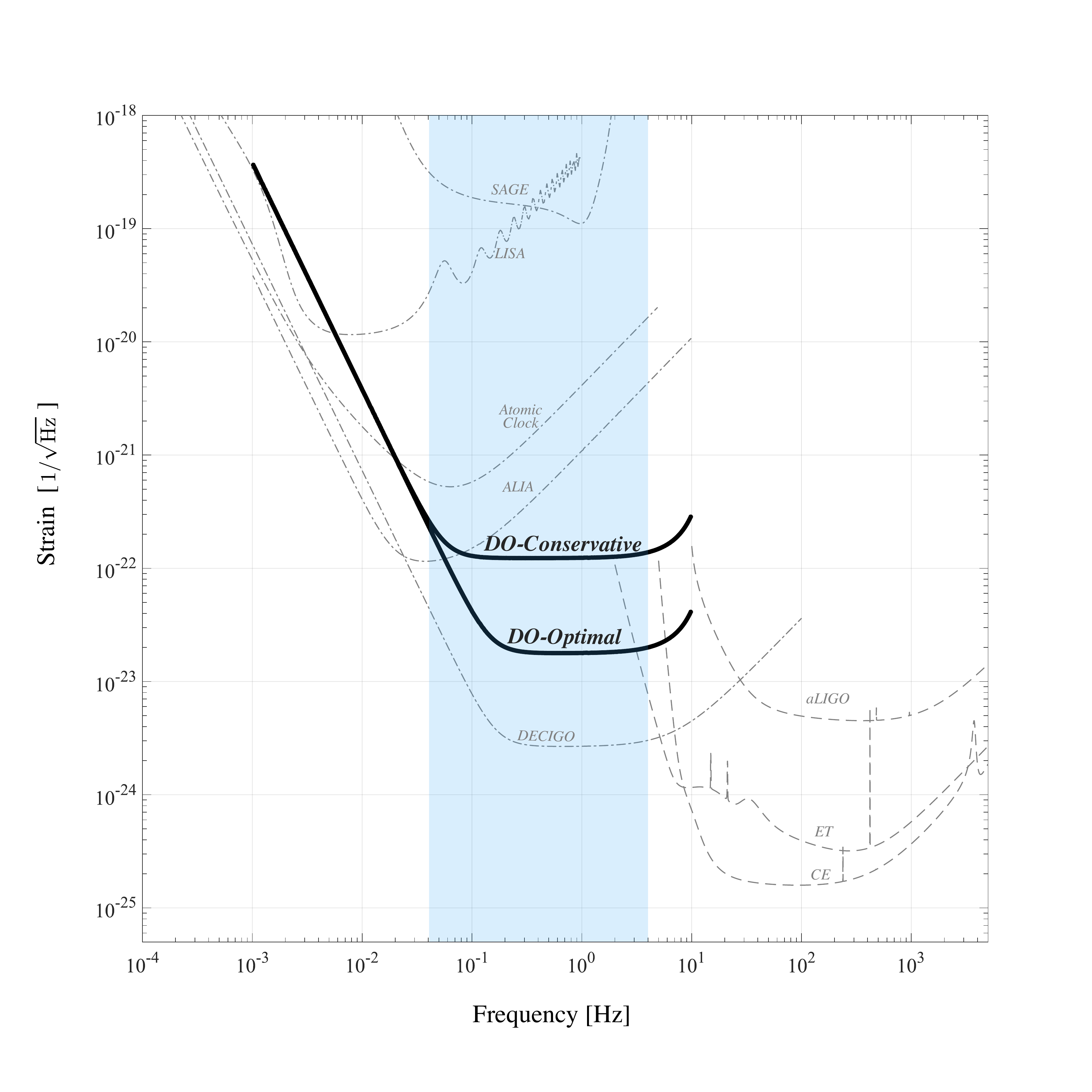} 
\caption{ 
Concept designs for Decihertz Observatories (DOs) fill the gap between \textit{LISA}~\cite{Audley:2017drz} and ground-based detectors like Advanced LIGO (aLIGO)~\cite{TheLIGOScientific:2014jea}, Cosmic Explorer (CE)~\cite{Evans:2016mbw} and the Einstein Telescope (ET)~\cite{Hild:2010id}. 
The example DO concepts \textit{SAGE}~\cite{Lacour:2018nws,Tino:2019tkb}, 
Atomic Clock~\cite{Kolkowitz:2016wyg,Sedda:2019uro}, 
\textit{ALIA}~\cite{Bender:2013nsa,Baker:2019pnp}, 
DO-Conservative, DO-Optimal~\cite{Sedda:2019uro-v1,Sedda:2019uro} and \textit{DECIGO}~\cite{Sato:2017dkf,Kawamura:2020pcg} span a diverse set of technologies and illustrate the potential range in sensitivities. 
}
\label{Fig1}
\end{figure}

The scientific return of a DO will depend upon its design. 
There are multiple potential technologies and mission concepts for observing the $0.01$--$1~\mathrm{Hz}$ GW spectrum. 
The \textit{Advanced Laser Interferometer Antenna} (\textit{ALIA})~\cite{Bender:2013nsa,Baker:2019pnp} is a heliocentric mission concept more sensitive than \textit{LISA} in the $0.1$--$1~\mathrm{Hz}$ range.
Other heliocentric DO concepts are \textit{Taiji}~\cite{Hu:2017mde,Guo:2018npi}, most sensitive around $0.01~\mathrm{Hz}$, and \textit{TianGo}~\cite{Kuns:2019upi}, most sensitive in the $0.1$--$10~\mathrm{Hz}$ range.
\textit{TianQin}~\cite{Luo:2015ght} is a Chinese geocentric mission concept.
The \textit{DECi-hertz Interferometer Gravitational-wave Observatory} (\textit{DECIGO})~\cite{Sato:2017dkf,Kawamura:2020pcg} is a more ambitious concept with $1000~\mathrm{km}$ Fabry--Perot cavity arms in heliocentric orbit; its precursor \textit{B-DECIGO} would be a $100~\mathrm{km}$ triangular interferometer in a geocentric orbit. 
The \textit{Big Bang Observer} (\textit{BBO}) is a concept consisting of four \textit{LISA} detectors in heliocentric orbits with combined peak sensitivity over $0.1$--$1~\mathrm{Hz}$ range~\cite{Crowder:2005nr}.  
More modest designs are the \textit{Geostationary Antenna for Disturbance-Free Laser Interferometry} (\textit{GADFLI})~\cite{McWilliams:2011ra} and \textit{geosynchronous Laser Interferometer Space Antenna} (\textit{gLISA})~\cite{Tinto:2011nr,Tinto:2014eua} which are geocentric concepts. 
The \textit{SagnAc interferometer for Gravitational wavE} (\textit{SAGE})~\cite{Lacour:2018nws,Tino:2019tkb} consists of three identical CubeSats in geosynchronous orbit. 
These concepts are mostly variations on the classic \textit{LISA} design of a laser interferometer.
In addition to laser interferometry, atomic-clock-based~\cite{Kolkowitz:2016wyg,Su:2017kng} and atom interferometer concepts are in development; for example, the \textit{Mid-band Atomic Gravitational Wave Interferometric Sensor} (\textit{MAGIS})~\cite{Graham:2017pmn} and the \textit{Atomic Experiment for Dark Matter and Gravity Exploration in Space} (\textit{AEDGE})~\cite{Bertoldi:2019tck} designs use atom interferometry. 
The range of technologies available mean that there are multiple possibilities for obtaining the necessary sensitivity in the decihertz range. 
Two illustrative \textit{LISA}-like designs, the more ambitious DO-Optimal and the less challenging DO-Conservative, are presented in \cite{Sedda:2019uro} to assess the potential range of science achievable with DOs.  
Potential sensitivity of DOs are illustrated in Figure~\ref{Fig1} in comparison to other gravitational-wave observatories.

\section{Summary}

Observing GWs in the decihertz range presents huge opportunities for advancing our understanding of both astrophysics and fundamental physics. 
The only prospect for decihertz observations is a space-based DO. 
Realising the rewards of these observations will require development of new detectors beyond \textit{LISA}. 
\emph{There are many challenges in meeting the requirements of DO concepts; however, there are also many promising technologies that could be developed to meet these goals}. 
A DO mission ready for launch in 2035--2050 is achievable, and the science payoff is worth the experimental effort.

\begin{acknowledgements}
{This summary is derived from a White Paper submitted 4 August 2019 to ESA's Voyage 2050 planning cycle on behalf of the \textit{LISA} Consortium 2050 Task Force~\cite{Sedda:2019uro-v1}}. 
Further space-based GW observatories considered by the \textit{LISA} Consortium 2050 Task Force include a microhertz observatory $\mu$\textit{Ares}~\cite{Sesana:2019vho}; a more sensitive millihertz observatory, the \textit{Advanced Millihertz
Gravitational-wave Observatory} (\textit{AMIGO})~\cite{Baibhav:2019rsa}, and a high angular-resolution observatory consisting of multiple DOs~\cite{Baker:2019ync}.

The authors thanks Pete Bender for insightful comments, and Adam Burrows and David Vartanyan for further suggestions.
MAS acknowledges financial support from the Alexander von Humboldt Foundation and the Deutsche Forschungsgemeinschaft (DFG, German Research Foundation) -- Project-ID 138713538 -- SFB 881 (``The Milky Way System''). 
CPLB is supported by the CIERA Board of Visitors Research Professorship. 
PAS acknowledges support from the Ram{\'o}n y Cajal Programme of the Ministry
of Economy, Industry and Competitiveness of Spain, as well as the COST Action
GWverse CA16104. This work was supported by the National Key R\&D Program of
China (2016YFA0400702) and the National Science Foundation of China (11721303). 
TB is supported by The Royal Society (grant URF\textbackslash R1\textbackslash 180009). 
EB is supported by National Science Foundation (NSF) Grants No.\ PHY-1912550 and AST-1841358, NASA ATP Grants No.\ 17-ATP17-0225 and 19-ATP19-0051, NSF-XSEDE Grant No.\ PHY-090003, and by the Amaldi Research Center, funded by the MIUR program ``Dipartimento di Eccellenza''~(CUP: B81I18001170001). This work has received funding from the European Union’s Horizon 2020 research and innovation programme under the Marie Skłodowska-Curie grant agreement No.\ 690904. 
DD acknowledges financial support via the Emmy Noether Research Group funded by the German Research Foundation (DFG) under grant no.\ DO 1771/1-1 and the Eliteprogramme for Postdocs funded by the Baden-Wurttemberg Stiftung.
JME is supported by NASA through the NASA Hubble Fellowship grant HST-HF2-51435.001-A awarded by the Space Telescope Science Institute, which is operated by the Association of Universities for Research in Astronomy, Inc., for NASA, under contract NAS5-26555. 
MLK acknowledges support from the NSF under grant DGE-0948017 and the Chateaubriand Fellowship from the Office for Science \& Technology of the Embassy of France in the United States. 
GN is partly supported by the ROMFORSK grant Project No.\ 302640 ``Gravitational Wave Signals From Early Universe Phase Transitions''.
IP acknowledges funding by Society in Science, The Branco Weiss Fellowship, administered by the ETH Zurich. 
AS is supported by the European Union's H2020 ERC Consolidator Grant ``Binary massive black hole astrophysics'' (grant agreement no.\ 818691 -- B Massive). 
LS was supported by the National Natural Science Foundation of China (11975027, 11991053, 11721303), the Young Elite Scientists Sponsorship Program by the China Association for Science and Technology (2018QNRC001), and the Max Planck Partner Group Program funded by the Max Planck Society.
NW is supported by a Royal Society--Science Foundation Ireland University Research Fellowship (grant UF160093). 
\end{acknowledgements}

\bibliographystyle{spphys}
\bibliography{biblio}

\end{document}